\documentclass{PoS}
\usepackage[utf8x]{inputenc}
\usepackage{amssymb}
\usepackage{amsmath}

\newcommand{\rt}{{\mathbf{r}}}
\newcommand{\xt}{{\mathbf{x}}}
\newcommand{\yt}{{\mathbf{y}}}

\newcommand{\bt}{{\mathbf{b}}}
\newcommand{\bti}{{\mathbf{b}_{i}}}
\newcommand{\gev}{\ \textrm{GeV}}
\newcommand{\der}{\mathrm{d}}
\newcommand{\xpom}{{x_\mathbb{P}}}
\newcommand{\A}{\mathcal{A}}

\title{Proton structure fluctuations: from HERA to the LHC}

\ShortTitle{Proton structure fluctuations: from HERA to the LHC}

\author{\speaker{Heikki Mäntysaari}\\
       Physics Department, Brookhaven National Laboratory, Upton, NY 11973, USA\\
        E-mail: \email{mantysaari@bnl.gov}}

\author{Björn Schenke\\
        Physics Department, Brookhaven National Laboratory, Upton, NY 11973, USA\\
        E-mail: \email{bschenke@bnl.gov}}
        
        \author{Chun Shen\\
        Physics Department, Brookhaven National Laboratory, Upton, NY 11973, USA\\
        E-mail: \email{cshen@bnl.gov}}
        
        \author{Prithwish Tribedy\\
        Physics Department, Brookhaven National Laboratory, Upton, NY 11973, USA\\
        Variable Energy Cyclotron Centre, HBNI, 1/AF, Bidhan Nagar, Kolkata 700064, India\\
        E-mail: \email{ptribedy@bnl.gov}}

\abstract{We constrain the amount of event-by-event fluctuations of the proton by comparing with the HERA coherent and incoherent diffractive $J/\Psi$ production data. We find that the HERA measurements prefer large geometric fluctuations. We then include the constrained fluctuating proton shapes in viscous hydrodynamical simulations of the proton-nucleus collisions at $\sqrt{s_{NN}}=5.02$ TeV, and find a good description of the mean transverse momentum and flow harmonics measured at the LHC.}

\FullConference{XXV International Workshop on Deep-Inelastic Scattering and Related Subjects\\
		3-7 April 2017\\
		University of Birmingham, UK}

\begin{document}

\section{Introduction}

Measurements of high-multiplicity proton-nucleus collisions at the LHC have revealed collective phenomena that have been traditionally interpreted as evidence of a creation of a hydrodynamically evolving fluid, the quark-gluon plasma. These signals include, for example, long range rapidity correlations and harmonic flow coefficients. For a review of recent collectivity measurements, the reader is referred to Ref.~\cite{Dusling:2015gta}. This has raised a natural question whether a hydrodynamically evolving medium is  created in proton-nucleus and even in high-multiplicity proton-proton collisions.
In Ref.~\cite{Mantysaari:2017cni} we quantitatively study if the IP-Glasma framework that has been  successful in describing a vast variety of LHC and RHIC heavy ion data (see e.g. Ref~\cite{Schenke:2012wb}) can be applied to high-multiplicity pA collisions.

In hydrodynamical simulations the initial state geometric anisotropies are converted into momentum space correlations and are observed e.g. as an elliptic ($v_2$) and triangular flow ($v_3$). Even tough the proton is on average round, the eccentricity of the initial state can be large if the proton geometry has large event-by-event fluctuations. Without these fluctuations, the explanation of the large elliptic and triangular flow in proton-nucleus collisions is difficult~\cite{Schenke:2014zha}. However, the inclusion of additional parameters that determine a fluctuating shape of the proton requires additional input besides measurements in proton-nucleus collisions in order to consistently test the quantitative agreement between the hydrodynamical picture and experimental data.

One possibility to constrain the proton structure fluctuations is given by exclusive vector meson (in this work $J/\Psi$) production measured at the HERA electron-proton collider. Unlike other DIS observables, diffractive scattering processes are sensitive to the transverse geometry of the target, as the impact parameter is a Fourier conjugate to the momentum transfer. These processes can be divided into two categories. In \emph{coherent diffraction}, where the target hadron remains in the same quantum state, the transverse momentum spectra of the produced vector mesons are directly related to the average density profile of the target~\cite{Miettinen:1978jb,Kowalski:2006hc}. On the other hand, in \emph{incoherent diffraction}, where the target breaks up but there is still no net color charge exchanged between the vector meson and the target, the cross section is proportional to the \emph{amount of fluctuations} of hte density profile~\cite{Miettinen:1978jb,Frankfurt:2008vi,Lappi:2010dd}. 

In this work, we constrain the proton shape fluctuations using  HERA diffractive $J/\Psi$ production data, and use the obtained fluctuating protons as an input for the hydrodynamical simulations of the proton-nucleus collisions at the LHC energy $\sqrt{s_{NN}}=5.02\,\mathrm{TeV}$. The diffractive vector meson production calculations are published in Refs.~\cite{Mantysaari:2016ykx,Mantysaari:2016jaz} and the hydrodynamical simulations in Ref.~\cite{Mantysaari:2017cni}.

\section{Theoretical framework: the IP-Glasma model}
\label{sec:fluctuations}
Instead of  a rotationally symmetric proton, we assume that the color charge density in the proton is distributed around the three hot spots whose positions in the transverse plane are sampled from a Gaussian distribution. The width of that distribution, $B_{qc}$, and the width of the density profile of each hot spots $B_q$, are free parameters to be determined by the diffractive DIS data. The density around each hot spot is parametrized as
$
T_q(\bt) = \frac{1}{2\pi B_q} e^{-\bt^2/(2B_q)}.
$

Writing the density profile of the proton as $T(\bt) = \frac{1}{3}\sum_{i=1}^3 T_q(\bt - \bti)$ we  use the IPsat dipole model fitted to the HERA structure function data~\cite{Rezaeian:2012ji} to calculate the saturation scale $Q_s$ at each point on the transverse plane. Then, solving the classical Yang-Mills equations we obtain the Wilson lines $V(\xt)$, which completely determine the initial stage. When we calculate diffractive vector meson production, for which the dipole-proton scattering amplitude is obtained as $N(\rt=\xt-\yt) = \mathrm{tr}(1 - V(\xt)V^\dagger(\yt))/N_c$. For more details, see Refs.~\cite{Mantysaari:2016ykx,Mantysaari:2016jaz,Mantysaari:2017cni}.

The scattering amplitude for  diffractive vector meson production 
can be written as~\cite{Kowalski:2006hc}
\begin{equation}
\label{eq:diff_amp}
 \A^{\gamma^* p \to V p}_{T,L}(\xpom,Q^2, \boldsymbol{\Delta}) = 2i\int \der^2 \rt \int \der^2 \bt \int \frac{\der z}{4\pi}  (\Psi^*\Psi_V)_{T,L}(Q^2, \rt,z)  e^{-i[\bt - (1-z)\rt]\cdot \boldsymbol{\Delta}}  N(\bt,\rt,\xpom).
\end{equation}
Here $Q^2$ is the virtuality of the photon, $\bt$ impact parameter, $\rt$ size of the quark-antiquark dipole and $z$ longitudinal momentum fraction of the photon carried by the quark.

The coherent  cross section is obtained by averaging the scattering amplitude over different configurations of the proton, and then taking the square:
\begin{equation}
\label{eq:coherent}
\frac{\der \sigma^{\gamma^* p \to V p}}{\der t} = \frac{1}{16\pi} \left| \langle \A^{\gamma^* p \to V p}(\xpom,Q^2,\boldsymbol{\Delta}) \rangle \right|^2.
\end{equation}
On the other hand, the incoherent cross section can be written as a variance \cite{Miettinen:1978jb} (see also Refs.~\cite{Frankfurt:2008vi,Lappi:2010dd}): 
\begin{equation}\label{eq:incoherent}
\frac{\der \sigma^{\gamma^* p \to V p^*}}{\der t}  = \frac{1}{16\pi} \left( \left\langle \left| \A^{\gamma^* p \to V p}(\xpom,Q^2,\boldsymbol{\Delta})  \right|^2 \right\rangle - \left| \langle  \A^{\gamma^* p \to V p}(\xpom,Q^2,\boldsymbol{\Delta}) \rangle \right|^2 \right)\,.
\end{equation}  

As the incoherent cross section is proportional to the variance of the amplitude \eqref{eq:diff_amp}, it measures the amount fluctuations of the proton density profile. On the other hand, the coherent cross section is sensitive to the average density profile, as the target average results in a dependency only on the average proton shape. Thus, requiring simultaneous description of HERA coherent and incoherent diffractive $J/\Psi$ production data~\cite{Alexa:2013xxa} allows us to constrain the unknown parameters describing the proton shape fluctuations. In Refs.~\cite{Mantysaari:2016ykx,Mantysaari:2016jaz} we find $B_{qc}=3.0$ GeV$^{-2}$ and $B_q=0.3$ GeV$^{-2}$. In addition, for the infrared cutoff parameter that is suppressing long distance Coulomb tails, we find $m=0.4$ GeV. 

 \begin{figure}[tb]
    \centering
    \begin{minipage}{.48\textwidth}
        \centering
        \includegraphics[width=0.8\textwidth]{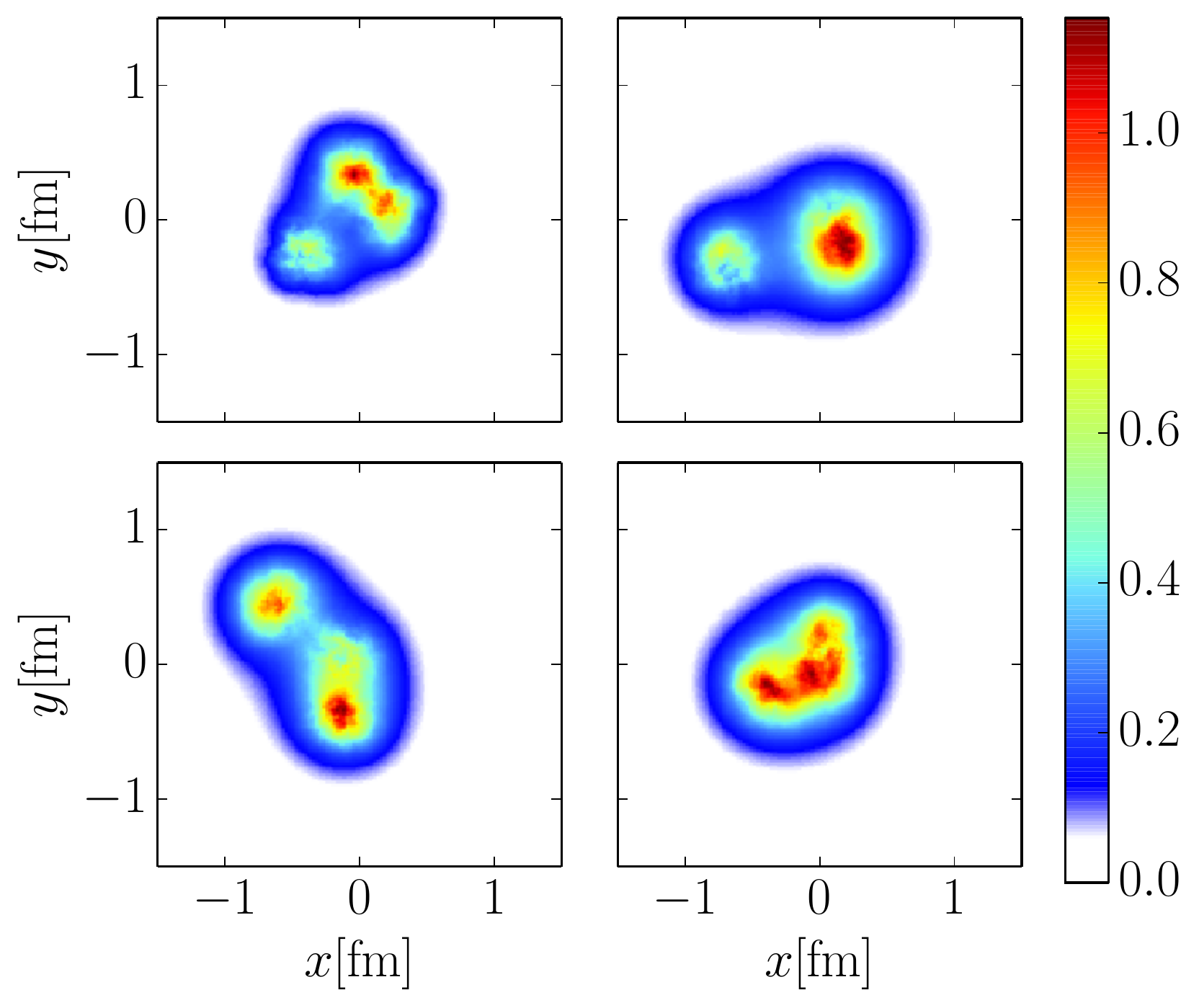} 
				\caption{Illustration of the event-by-event fluctuations of the proton density profile.}
		\label{fig:ipglasma}
    \end{minipage} \quad
    \begin{minipage}{0.48\textwidth}
        \centering
      \includegraphics[width=\textwidth]{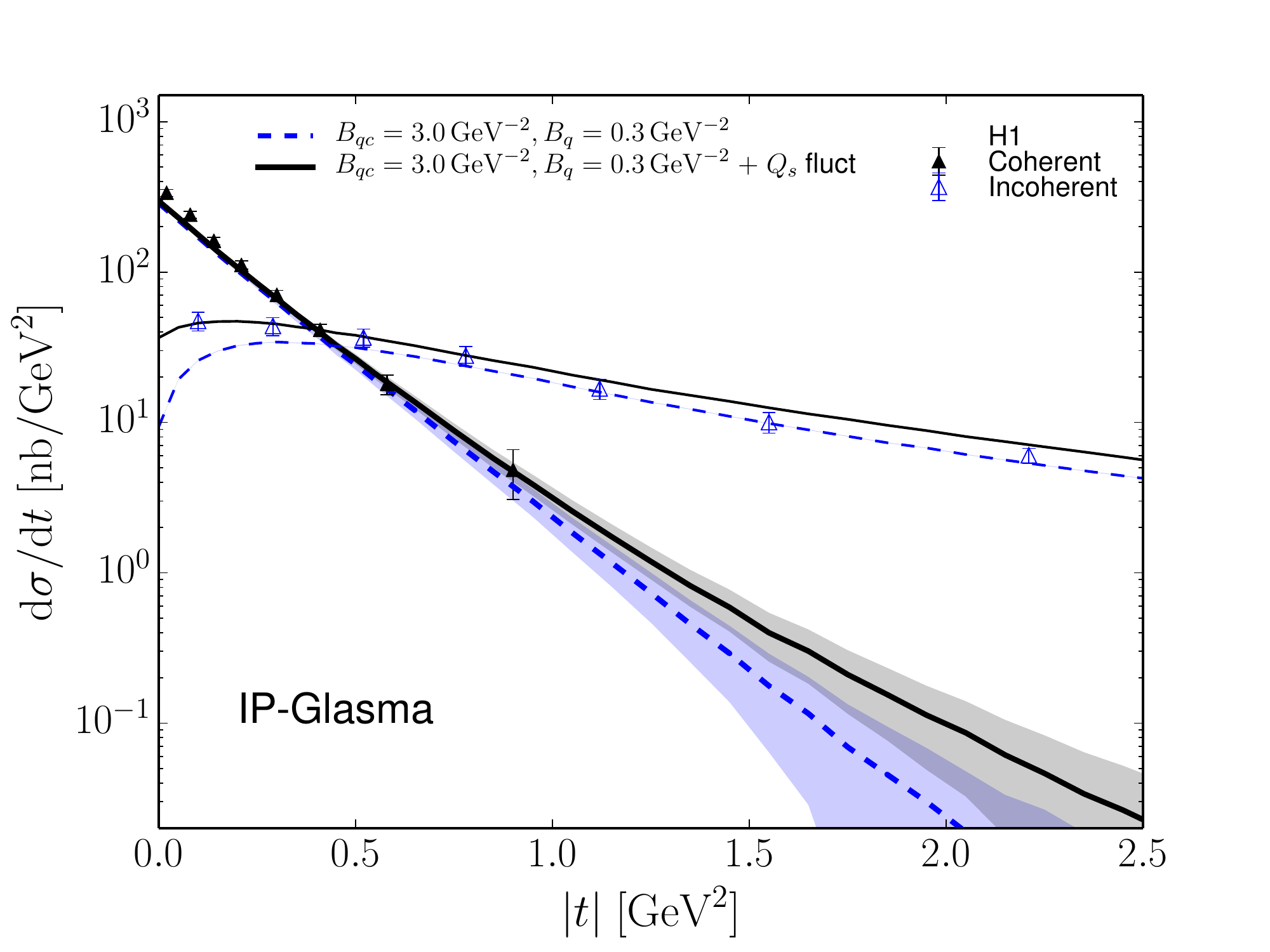} 
				\caption{Coherent and incoherent diffractive $J/\Psi$ production compared with the H1 data~\cite{Alexa:2013xxa}.}
		\label{fig:diffractive_spectra}
    \end{minipage}
\end{figure}

Proton-nucleus collisions are described as follows.
We simulate the hydrodynamical evolution of the produced matter   
using {\sc Music}~\cite{Schenke:2010nt,Schenke:2010rr}. We use the same second order transport parameters as in Ref.~\cite{Ryu:2015vwa}, and use both temperature dependent and constant shear viscosity $\eta/s=0.2$. We also include bulk viscosity and use a temperature dependent $\zeta/s$ as in~\cite{Ryu:2015vwa}, which is essential in small systems to obtain average transverse momenta compatible with the LHC data. Before $\tau_0=0.2\,\mathrm{fm}/c$, when the hydrodynamical evolution is started, the pre-thermal evolution is described by the solutions of the classical Yang-Mills equations as implemented in the IP-Glasma model.  The full energy-momentum tensor $T^{\mu\nu}$ at $\tau_0$ is matched to the hydrodynamical phase including the initial shear stress tensor and bulk viscous pressure. 
The hydrodynamic equations are evolved with a Lattice Equation of State s95p-v1~\cite{Huovinen:2009yb}. The switching temperature is set to $T_\text{switch}=155$ MeV, after which particles are sampled using the Cooper-Frye prescription and are allowed to propagate in the hadronic cascade model UrQMD~\cite{Bleicher:1999xi}. For more details, see Ref.~\cite{Mantysaari:2017cni}.

\section{Results}
The cross section for the coherent and incoherent diffractive $J/\Psi$ production at $W=75\gev$, after  fitting the parameters controlling the shape of the fluctuating proton, $B_q, B_{qc}$ and $m$,  is shown in Fig.~\ref{fig:diffractive_spectra}. We note that without geometrical fluctuations the experimentally measured incoherent cross section~\cite{Alexa:2013xxa} would be underestimated by roughly two orders of magnitude, see Ref.~\cite{Mantysaari:2016jaz}.
The obtained proton profiles (at $x \approx 10^{-3}$) are illustrated in Fig.~\ref{fig:ipglasma}. In Fig.~\ref{fig:diffractive_spectra} we also show the effect of including, on top of the geometric fluctuations, additional fluctuations of the saturation scale (see~\cite{Mantysaari:2016jaz}). As these fluctuations take place at longer distance scale, their effect is largest at small $|t|$ where these additional fluctuations make the agreement with the H1 data better.

\begin{figure}[tb]
    \centering
    \begin{minipage}{.48\textwidth}
        \centering
        \includegraphics[width=\textwidth]{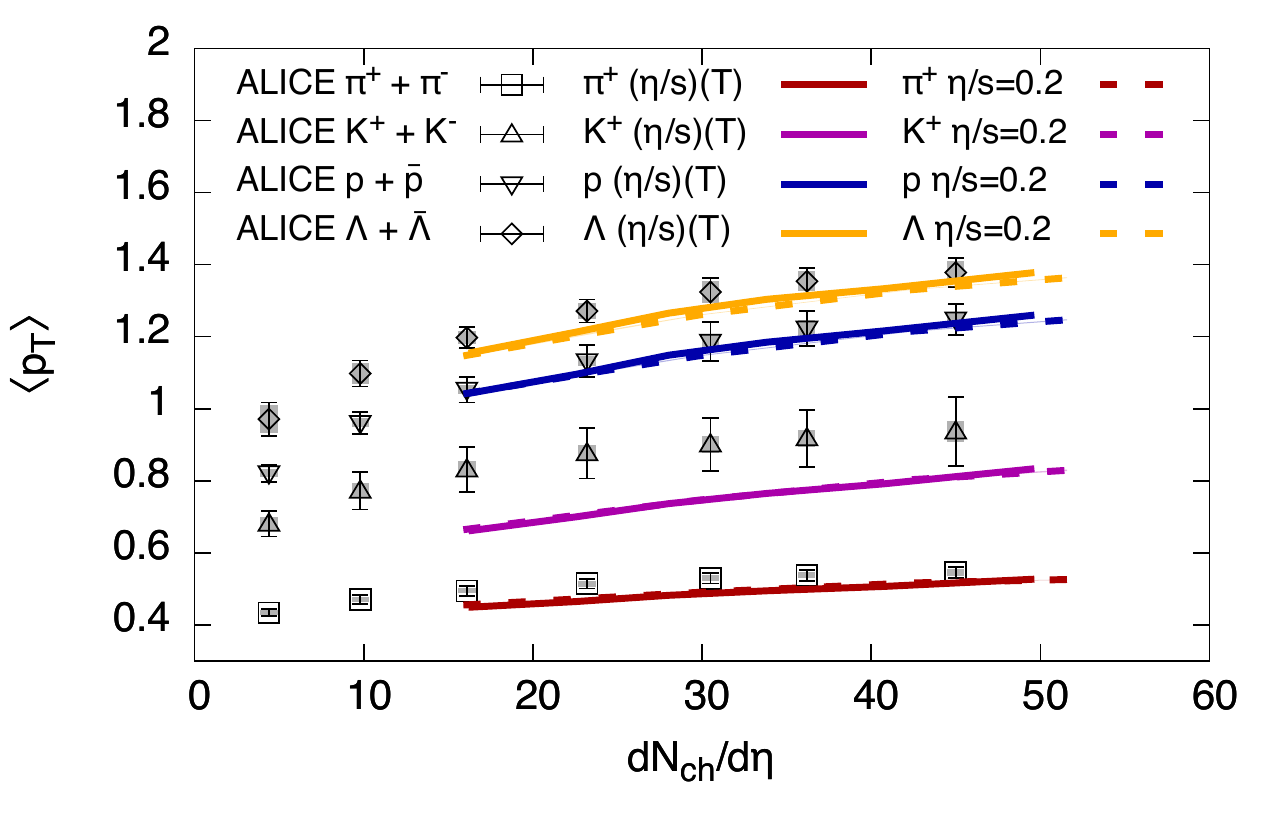} 
				\caption{The average transverse momentum of identified particles as a function of multiplicity compared with the ALICE data~\cite{Abelev:2013haa}.}.
		\label{fig:meanpt}
    \end{minipage} \quad
    \begin{minipage}{0.48\textwidth}
        \centering
      \includegraphics[width=\textwidth]{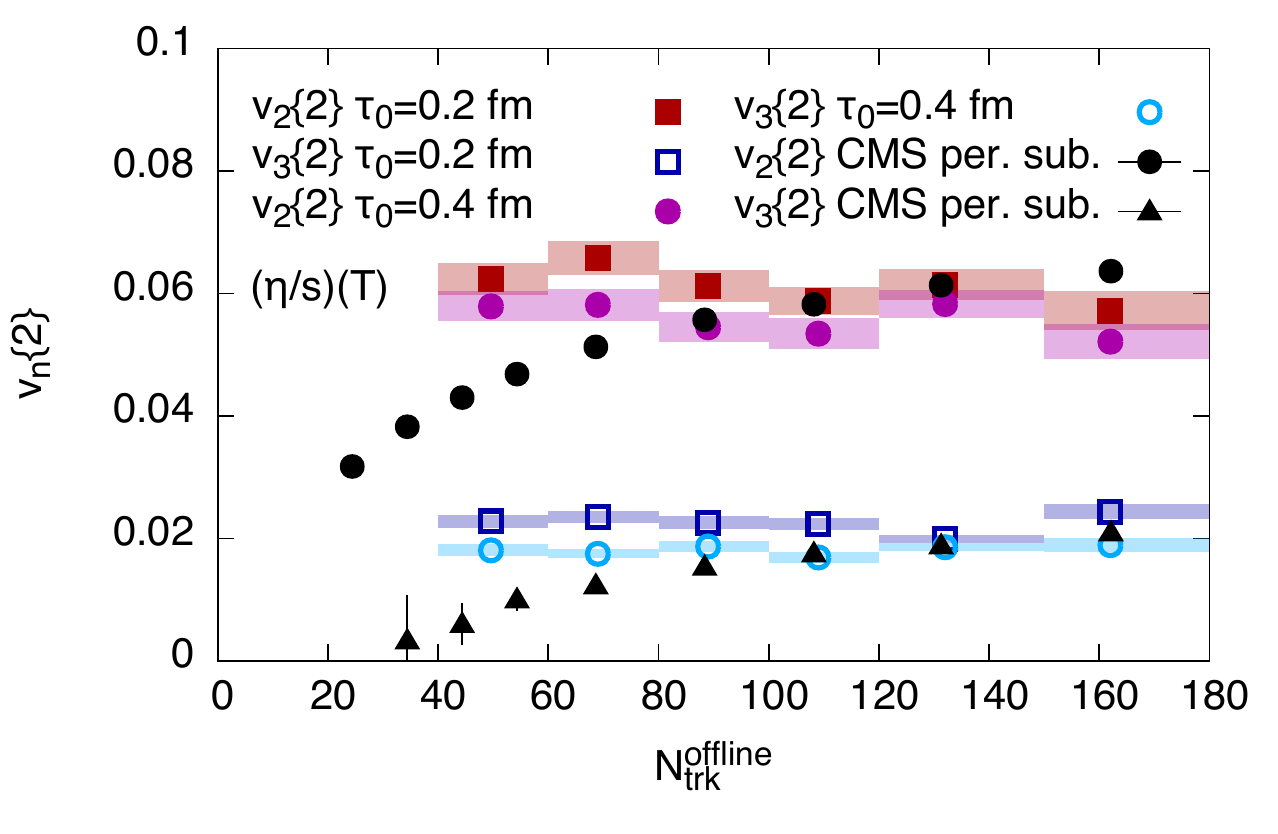} 
				\caption{Elliptic ($v_2$) and triangular ($v_3$) flow in p+Pb collisions at $\sqrt{s_{NN}}=5.02$ TeV as a function of multiplicity compared with the CMS data~ \cite{Chatrchyan:2013nka}.}
		\label{fig:v2v3}
    \end{minipage}
\end{figure}

We then use the same fluctuating proton structure to do hydrodynamical simulations of the proton-nucleus collisions. In Fig.~\ref{fig:meanpt} we show the mean transverse momentum of identified hadrons compared with the ALICE data~\cite{Abelev:2013haa}. Overall we find a good agreement with the measurements, except in case of charged kaons. The results are also unchanged when a constant $\eta/s$ is replaced by a temperature dependent parametrization.

The elliptic and triangular flow coefficients $v_2$ and $v_3$ at $\sqrt{s_{NN}}=5.02$ TeV are compared with the ALICE data in Fig.~\ref{fig:v2v3}. We note that in case of no geometric fluctuations in the proton, the data would be significantly underestimated due to the small initial state eccentricity~\cite{Schenke:2014zha}. With the geometric fluctuations constrained by the HERA data, we find a good description with both $v_2$ and $v_3$ data. For more details, see Ref.~\cite{Mantysaari:2017cni}.

\section{Conclusions}
The amount of geometric fluctuations  in the proton wave function is constrained using the HERA diffractive $J/\Psi$ production data. Large event-by-event geometric fluctuations are found to be necessary in order to describe the experimentally measured incoherent cross section, which is sensitive to the variance of the density profile. 
When simulating proton-nucleus collisions with relativistic hydrodynamics using an initial state determined by these fluctuating protons, we find a good description of the flow harmonics in high multiplicity events. 
In Ref.~\cite{Mantysaari:2017dwh} the necessity of having large geometric fluctuations was also found to improve the description of the exclusive $J/\Psi$ production data in ultraperipheral heavy ion collisions at the LHC.

\section*{Acknowledgments}
This work was supported under DOE Contract No. DE-SC0012704 and used resources of the National Energy Research Scientific Computing Center, supported by the Office of Science of the U.S. Department of Energy under Contract No. DE-AC02-05CH11231. BPS acknowledges a DOE Office of Science Early Career Award.
\bibliographystyle{JHEP}
\bibliography{../../refs}

\providecommand{\href}[2]{#2}\begingroup\raggedright\begin{thebibliography}{10}

\bibitem{Dusling:2015gta}
K.~Dusling, W.~Li and B.~Schenke, \emph{{Novel collective phenomena in
  high-energy proton–proton and proton–nucleus collisions}},
  \href{http://dx.doi.org/10.1142/S0218301316300022}{\emph{Int. J. Mod. Phys.}
  {\bfseries E25} (2016) 1630002},
  [\href{https://arxiv.org/abs/1509.07939}{{\ttfamily 1509.07939}}].

\bibitem{Mantysaari:2017cni}
H.~Mäntysaari, B.~Schenke, C.~Shen and P.~Tribedy, \emph{{Imprints of
  fluctuating proton shapes on flow in proton-lead collisions at the LHC}},
  \href{https://arxiv.org/abs/1705.03177}{{\ttfamily 1705.03177}}.

\bibitem{Schenke:2012wb}
B.~Schenke, P.~Tribedy and R.~Venugopalan, \emph{{Fluctuating Glasma initial
  conditions and flow in heavy ion collisions}},
  \href{http://dx.doi.org/10.1103/PhysRevLett.108.252301}{\emph{Phys. Rev.
  Lett.} {\bfseries 108} (2012) 252301},
  [\href{https://arxiv.org/abs/1202.6646}{{\ttfamily 1202.6646}}].

\bibitem{Schenke:2014zha}
B.~Schenke and R.~Venugopalan, \emph{{Eccentric protons? Sensitivity of flow to
  system size and shape in p+p, p+Pb and Pb+Pb collisions}},
  \href{http://dx.doi.org/10.1103/PhysRevLett.113.102301}{\emph{Phys. Rev.
  Lett.} {\bfseries 113} (2014) 102301},
  [\href{https://arxiv.org/abs/1405.3605}{{\ttfamily 1405.3605}}].

\bibitem{Miettinen:1978jb}
H.~I. Miettinen and J.~Pumplin, \emph{{Diffraction Scattering and the Parton
  Structure of Hadrons}},
  \href{http://dx.doi.org/10.1103/PhysRevD.18.1696}{\emph{Phys. Rev.}
  {\bfseries D18} (1978) 1696}.

\bibitem{Kowalski:2006hc}
H.~Kowalski, L.~Motyka and G.~Watt, \emph{{Exclusive diffractive processes at
  HERA within the dipole picture}},
  \href{http://dx.doi.org/10.1103/PhysRevD.74.074016}{\emph{Phys. Rev.}
  {\bfseries D74} (2006) 074016},
  [\href{https://arxiv.org/abs/hep-ph/0606272}{{\ttfamily hep-ph/0606272}}].

\bibitem{Frankfurt:2008vi}
L.~Frankfurt, M.~Strikman, D.~Treleani and C.~Weiss, \emph{{Evidence for color
  fluctuations in the nucleon in high-energy scattering}},
  \href{http://dx.doi.org/10.1103/PhysRevLett.101.202003}{\emph{Phys. Rev.
  Lett.} {\bfseries 101} (2008) 202003},
  [\href{https://arxiv.org/abs/0808.0182}{{\ttfamily 0808.0182}}].

\bibitem{Lappi:2010dd}
T.~Lappi and H.~M{\"a}ntysaari, \emph{{Incoherent diffractive
  $J/\Psi$-production in high energy nuclear DIS}},
  \href{http://dx.doi.org/10.1103/PhysRevC.83.065202}{\emph{Phys. Rev.}
  {\bfseries C83} (2011) 065202},
  [\href{https://arxiv.org/abs/1011.1988}{{\ttfamily 1011.1988}}].

\bibitem{Mantysaari:2016ykx}
H.~Mäntysaari and B.~Schenke, \emph{{Evidence of strong proton shape
  fluctuations from incoherent diffraction}},
  \href{http://dx.doi.org/10.1103/PhysRevLett.117.052301}{\emph{Phys. Rev.
  Lett.} {\bfseries 117} (2016) 052301},
  [\href{https://arxiv.org/abs/1603.04349}{{\ttfamily 1603.04349}}].

\bibitem{Mantysaari:2016jaz}
H.~Mäntysaari and B.~Schenke, \emph{{Revealing proton shape fluctuations with
  incoherent diffraction at high energy}},
  \href{http://dx.doi.org/10.1103/PhysRevD.94.034042}{\emph{Phys. Rev.}
  {\bfseries D94} (2016) 034042},
  [\href{https://arxiv.org/abs/1607.01711}{{\ttfamily 1607.01711}}].

\bibitem{Rezaeian:2012ji}
A.~H. Rezaeian, M.~Siddikov, M.~Van~de Klundert and R.~Venugopalan,
  \emph{{Analysis of combined HERA data in the Impact-Parameter dependent
  Saturation model}},
  \href{http://dx.doi.org/10.1103/PhysRevD.87.034002}{\emph{Phys. Rev.}
  {\bfseries D87} (2013) 034002},
  [\href{https://arxiv.org/abs/1212.2974}{{\ttfamily 1212.2974}}].

\bibitem{Alexa:2013xxa}
{\scshape H1} collaboration, C.~Alexa et~al., \emph{{Elastic and
  Proton-Dissociative Photoproduction of J$/\Psi$ Mesons at HERA}},
  \href{http://dx.doi.org/10.1140/epjc/s10052-013-2466-y}{\emph{Eur. Phys. J.}
  {\bfseries C73} (2013) 2466},
  [\href{https://arxiv.org/abs/1304.5162}{{\ttfamily 1304.5162}}].

\bibitem{Schenke:2010nt}
B.~Schenke, S.~Jeon and C.~Gale, \emph{{(3+1)D hydrodynamic simulation of
  relativistic heavy-ion collisions}},
  \href{http://dx.doi.org/10.1103/PhysRevC.82.014903}{\emph{Phys. Rev.}
  {\bfseries C82} (2010) 014903},
  [\href{https://arxiv.org/abs/1004.1408}{{\ttfamily 1004.1408}}].

\bibitem{Schenke:2010rr}
B.~Schenke, S.~Jeon and C.~Gale, \emph{{Elliptic and triangular flow in
  event-by-event (3+1)D viscous hydrodynamics}},
  \href{http://dx.doi.org/10.1103/PhysRevLett.106.042301}{\emph{Phys. Rev.
  Lett.} {\bfseries 106} (2011) 042301},
  [\href{https://arxiv.org/abs/1009.3244}{{\ttfamily 1009.3244}}].

\bibitem{Ryu:2015vwa}
S.~Ryu, J.~F. Paquet, C.~Shen, G.~S. Denicol, B.~Schenke, S.~Jeon et~al.,
  \emph{{Importance of the Bulk Viscosity of QCD in Ultrarelativistic Heavy-Ion
  Collisions}},
  \href{http://dx.doi.org/10.1103/PhysRevLett.115.132301}{\emph{Phys. Rev.
  Lett.} {\bfseries 115} (2015) 132301},
  [\href{https://arxiv.org/abs/1502.01675}{{\ttfamily 1502.01675}}].

\bibitem{Huovinen:2009yb}
P.~Huovinen and P.~Petreczky, \emph{{QCD Equation of State and Hadron Resonance
  Gas}}, \href{http://dx.doi.org/10.1016/j.nuclphysa.2010.02.015}{\emph{Nucl.
  Phys.} {\bfseries A837} (2010) 26--53},
  [\href{https://arxiv.org/abs/0912.2541}{{\ttfamily 0912.2541}}].

\bibitem{Bleicher:1999xi}
M.~Bleicher et~al., \emph{{Relativistic hadron hadron collisions in the
  ultrarelativistic quantum molecular dynamics model}},
  \href{http://dx.doi.org/10.1088/0954-3899/25/9/308}{\emph{J. Phys.}
  {\bfseries G25} (1999) 1859--1896},
  [\href{https://arxiv.org/abs/hep-ph/9909407}{{\ttfamily hep-ph/9909407}}].

\bibitem{Abelev:2013haa}
{\scshape ALICE} collaboration, B.~B. Abelev et~al., \emph{{Multiplicity
  Dependence of Pion, Kaon, Proton and Lambda Production in p-Pb Collisions at
  $\sqrt{s_{NN}}$ = 5.02 TeV}},
  \href{http://dx.doi.org/10.1016/j.physletb.2013.11.020}{\emph{Phys. Lett.}
  {\bfseries B728} (2014) 25--38},
  [\href{https://arxiv.org/abs/1307.6796}{{\ttfamily 1307.6796}}].

\bibitem{Chatrchyan:2013nka}
{\scshape CMS} collaboration, S.~Chatrchyan et~al., \emph{{Multiplicity and
  transverse momentum dependence of two- and four-particle correlations in pPb
  and PbPb collisions}},
  \href{http://dx.doi.org/10.1016/j.physletb.2013.06.028}{\emph{Phys. Lett.}
  {\bfseries B724} (2013) 213--240},
  [\href{https://arxiv.org/abs/1305.0609}{{\ttfamily 1305.0609}}].

\bibitem{Mantysaari:2017dwh}
H.~Mäntysaari and B.~Schenke, \emph{{Probing subnucleon scale fluctuations in
  ultraperipheral heavy ion collisions}},
  \href{https://arxiv.org/abs/1703.09256}{{\ttfamily 1703.09256}}.

\end{thebibliography}\endgroup
\end{document}